\documentclass[preprint]{jfm}
\usepackage{graphicx}
\usepackage{epsfig}
\usepackage{bm}
\usepackage{amsmath}
\usepackage{amssymb}
\usepackage{wasysym}
\usepackage{epstopdf}
\usepackage{color}
\usepackage{xcolor}
\usepackage{epstopdf}
\usepackage{upgreek}
\usepackage{subfloat}
\usepackage{graphicx}
\usepackage{dcolumn}
\usepackage{bm}
\usepackage{latexsym}
\usepackage{subfloat}
\usepackage[amssymb]{SIunits}
\usepackage{subfig}
\usepackage{psfrag}
\usepackage{xcolor}
\usepackage{natbib}
\usepackage{epstopdf}
\usepackage{amsmath}
\usepackage{comment}
\graphicspath{{Figures//}}

\usepackage[colorlinks=false, bookmarks=true]{hyperref}

\def\tb{\textcolor{black}}

\usepackage[applemac]{inputenc}	
\usepackage[english]{babel} 
\newcommand{\reff}[1]{\ref{fig:#1}} 
\newcommand{\refe}[1]{\eqref{eq:#1}} 

\newcommand{\pen}[0]{\hspace{-0.1em}} 
\newcommand{\nsd}[1]{\mbox{\textit{#1}}} 
\newcommand{\ca}{\nsd{Ca}} 
\newcommand{\dd}{\mathrm{d}} 
\newcommand{\ed}{\mathrm{e}}

\newcommand{\vect}[1]{\boldsymbol{#1}}

\title[Levitation of a drop over a moving surface]{Levitation of a drop over a moving surface} 
\author[Lhuissier, H., Tagawa, Y., Tran, T., and Sun, C.]
{Henri\ns Lhuissier$^{1}$, Yoshiyuki\ns Tagawa$^{1,2}$,\ns Tuan\ns Tran$^{1}$ and
Chao\ns Sun$^{1}$\ns}

\affiliation{$^1$Physics of Fluids Group, Faculty of Science and Technology, J.M. Burgers Center for Fluid Dynamics, University of Twente, PO Box 217, 7500 AE  Enschede, The Netherlands\\[\affilskip]
$^2$ Department of Mechanical Systems Engineering, Tokyo University of Agriculture and Technology, 2-24-16 Nakacho, Koganei-city, Tokyo, Japan\\[\affilskip]}

\date{\today}

\begin{document}
\maketitle


\begin{abstract}
We investigate the levitation of a drop gently deposited onto the inner wall of a rotating hollow cylinder. For a sufficient velocity of the wall, the drop steadily levitates over a thin air film and reaches a stable angular position in the cylinder, where the  drag and lift balance the weight of the drop. Interferometric measurement yields the three-dimensional (3D) air film thickness under the drop and reveals the asymmetry of the profile along the direction of the wall motion. A two-dimensional (2D) model is presented which explains the levitation mechanism, captures the main characteristics of the air film shape and predicts two asymptotic regimes for the film thickness $h_0$: For large drops $h_0\sim\ca^{2/3}\kappa_b^{-1}$, as in the Bretherton problem, where $\ca$ is the capillary number based on the air viscosity and $\kappa_b$ is the curvature at the bottom of the drop. For small drops $h_0\sim\ca^{4/5}(a\kappa_b)^{4/5}\kappa_b^{-1}$, where $a$ is the capillary length. 
\vspace{-4mm}
\end{abstract}

\section{Introduction}\label{sec:Introduction}

	A drop falling onto a solid or a 
	liquid surface, as a raindrop onto a wall or a puddle
	usually makes contact with the impacted surface.
	However, under certain conditions, the persistency of the gas film separating the drop from the surface prevents contact. This counter-intuitive phenomenon has been investigated, both for fundamental interests and for its importance in various applications, such as droplet combustion, separation of emulsions, and spray painting. Configurations promoting non-contact have been explored, e.g. encapsulating the drop by a hydrophobic powder \citep{Aussillous2001} or oscillating the liquid surface to periodically renew the air film \citep{Couder2005}.
	
	We consider here a different situation where the levitation is maintained by moving a surface, which 
	continously
	entrains 
	air under the drop that balances the drainage of the film. 
	This phenomenon was already 
	observed 
	in a slightly different situation in which 
	a drop levitates over a 
	hydraulic jump \citep{Sreenivas1999,Pirat2010}. \cite{Sreenivas1999} suggested the levitation is due to the pressure built-up generated by the lubrication flow under the drop. 
	\cite{Neitzel2002} considered a situation where the drop is not free to move: a drop is fixed on a solid 
	plate 
	and pressed against a moving solid surface. They measured the air film thickness and observed the asymmetric deformation of the drop. Later \cite{Smith06} also realized simulations of the flow using lubrication theory but did not give any prediction for the film thickness or air drag.

	In order to shed light on the mechanism of levitation and study the details of the air film, we present here a simple experimental system allowing a steady levitation configuration in which the drop is free to move and deform. It is inspired from a similar configuration, used to investigate the motion of a bubble or a solid sphere in a rotating flow \tb{\citep{Bluemink2005, Bluemink2008, Bluemink2010, Tagawa2013}}, and consists of a drop gently deposited onto the inner wall of a hollow solid cylinder rotating around its horizontal axis. For a sufficient rotation speed, the drop levitates and reaches a stable angular position, at which the drag and lift balance the drop's weight. This stable levitation suggests a steady flow settles around and underneath the drop. 
	It allows us to measure the air drag on the drop and the 3D shape of the 
	air film where the lift force is generated.
	

\vspace{-7mm}
\section{Experimental setup and observations}\label{sec:expeobs}


\begin{figure}
	\centerline{\includegraphics[width=.7\textwidth]{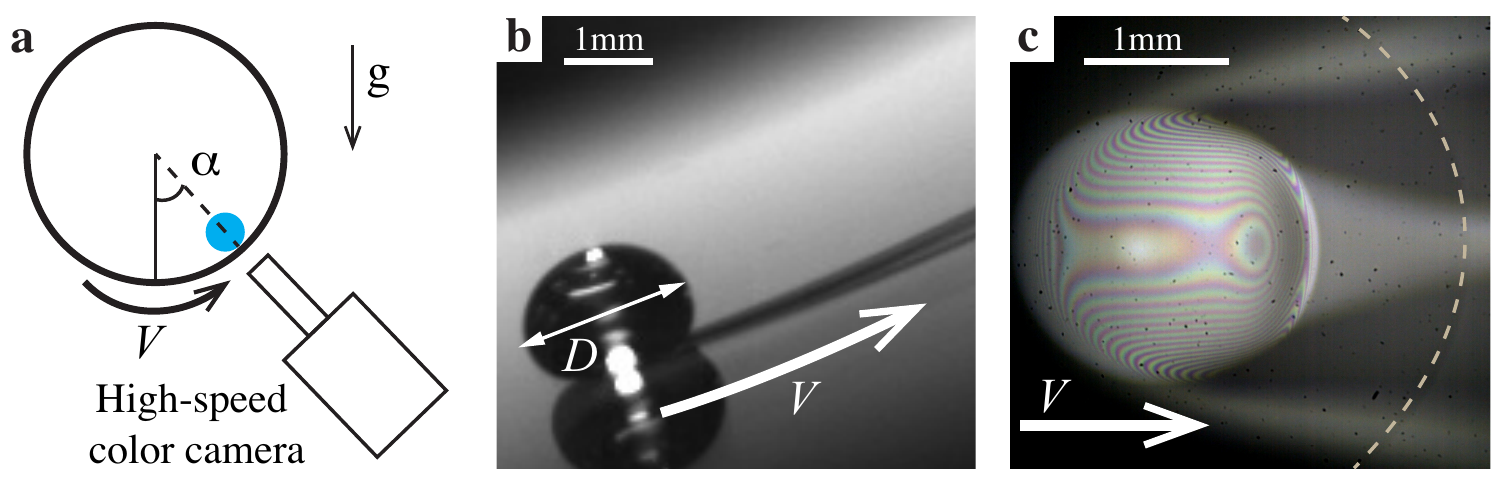}}
	\vspace{-3mm}
	\caption{\small{(a) Schematic (not to scale) of the experimental setup. 
	(b) Side view of a levitating droplet.
	(c) Interferometric measurements of the air layer under the droplet.
}}
	\label{setup}
\vspace{-5mm}
\end{figure}

	The experimental setup we used to study levitating drops is 
	shown
	in figure \ref{setup}(a). It consists of a hollow glass cylinder, with length 500\,mm and diameter 100\,mm, rotated about its axis in the horizontal direction, 
	with circumferential velocity up to 3.0$\,$m\,s$^{-1}$.
Drops of silicone oil (viscosity $\nu = 50\,$cSt, surface tension $\sigma=20.8\,$mN$\,$m$^{-1}$, and density $\rho = 961\,$kg$\,$m$^{-3}$) with diameter between 1$\,$mm and 3$\,$mm 
are gently deposited on the inner surface of the rotating cylinder. When the wall velocity $V$ is sufficiently large, the drop levitates and reaches a stable angular position $\alpha$. 
The inner surface of the cylinder is generally covered with a thin layer of the same liquid as in the drop, with thickness less than 1.0$\,$mm. This liquid layer, due to its smooth surface, significantly lengthens the levitating time. Nonetheless, it does not affect the levitation mechanism and the results presented below: for fixed wall velocity and drop size, 
we observed 
similar
equilibrium angular position of the levitating drop, whether the cylinder's surface is dry or wetted.
For each experiment, the motion of the drop, its apparent diameter $D$ and the angular position $\alpha$ (see figure \ref{setup}), are measured on side-view recordings, from a high-speed camera (SA1.1, Photron Inc.) which aims horizontally at the drop (not shown in figure \ref{setup}(a)). A typical side-view of a drop levitating over a thin liquid layer is shown in figure \ref{setup}(b).
	
\begin{figure}
	\centerline{\includegraphics[width=0.8\textwidth]{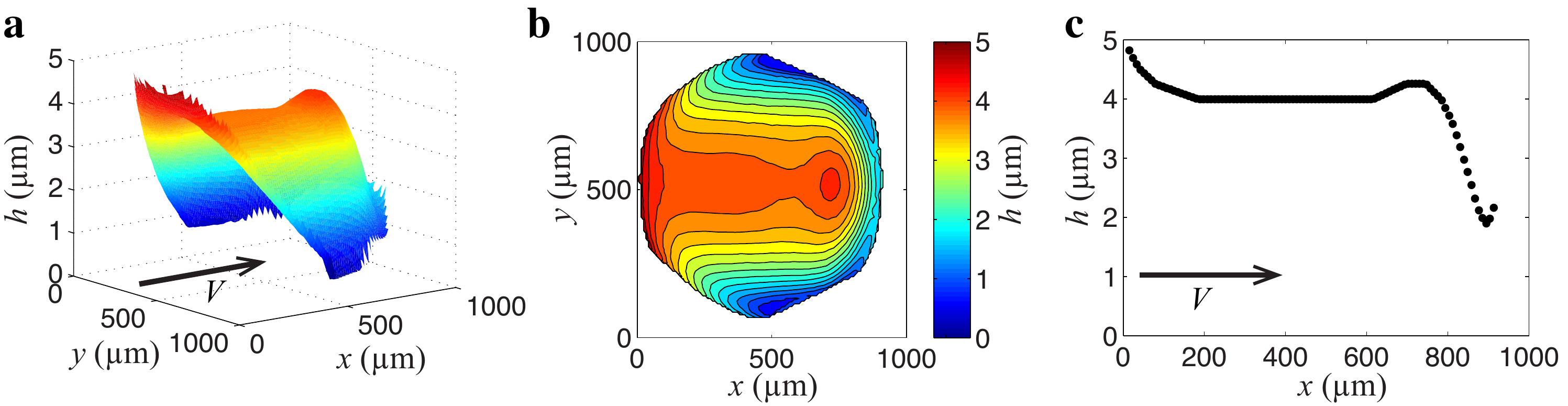}}
	\vspace{-3mm}
	\caption{\small{(a) 3D profile of the air film under the drop shown in figure \ref{setup}(c).
	 (b) Contour of the air film thickness profile.
	 (c) Thickness along the plane of symmetry of the air film ($y=500\,\mu$m). The arrows in (a) and (c) indicate
	 the direction of the wall motion.}}
	\label{3Dprofile}
\vspace{-5mm}
\end{figure}

	The levitation of a drop undoubtedly indicates the existence of a thin air film underneath the drop. To measure the thickness of this air film, we use a color interferometry method, which was described in details by \cite{veen12}, and successfully applied to measure the air layer thickness under droplets impacting on either solid or liquid surfaces \citep{veen12,tran12,tran13}. A color high-speed camera (SA2, Photron Inc.) is connected to a long working distance microscope and aims perpendicularly at the cylinder's wall, as shown in figure \ref{setup}(a). As white light is supplied through a coaxial port of the microscope and is reflected both from the bottom surface of the drop and from the top surface of the liquid layer, a colorful interference pattern is formed and recorded. A typical image of this pattern is shown in figure \ref{setup}(c). From a small strip across the fringes, we extract the absolute thickness of the air film \citep[see][]{veen12} and reconstruct the whole 3D air film profile by following the fringes with known thickness. In figure \ref{3Dprofile}(a) and (b), we show the 3D profile and the contour plot extracted from the interference pattern shown in figure \ref{setup}(c). 
	The thickness profile
	in the plane of symmetry of the air film 
	is shown in figure \ref{3Dprofile}(c).


	For each drop size $D$, levitation was only observed above a critical velocity. Below this velocity, the drop immediately touches the surface and 
	coalesces with
	the moving wall. Levitation is observed at lower velocity for small drops than for large ones. 
	For a fixed drop size $D$, the equilibrium angular position $\alpha$ increases with $V$, as the ratio of the air drag to the drop's weight increases.
	
	The viscosity of the drop and liquid layer also influences the levitation. We observed that levitation is facilitated for viscosities around 50\,cSt. For smaller viscosities, the drop may levitate for a short time but, due to the lower damping of capillary waves, the drop surface starts oscillating until it eventually touches the surface. For much larger viscosities, the drop does not levitate at all and makes contact immediately. 
	This may be due to the fact that 
	the relaxation time for the deformation of the drop becomes larger than the time for the drainage of the air film: the drop shape adapts too slowly to the flow under it to generate a sufficient lift before contact.  
	
	For $\nu = 50$\,cSt (the only viscosity we will consider from now on), levitating drops appear stationary and $\alpha$ changes very little with time. More surprisingly, the liquid in the drop was found to be essentially motionless. By seeding the drop with particles, we could measure the velocity of the drop inner motion to be of the order of 1\,mm\,s$^{-1}$, 
	that is to say negligible with respect to the wall velocity $V\sim 1\,$m s$^{-1}$. 
This suggests that the drop's rotational and translational motions do not contribute to the levitation mechanism, a conclusion we will use in \S\ref{sec:model}. 

	The interferometric measurements carry the information on the shape of the air film underneath the drop. As figure \ref{3Dprofile} reveals, the bottom of the drop is essentially flat: the air film is typically a few micrometers thick 
	and has an aspect ratio larger than 10$^2$. 
	It has a flat portion in the middle followed by a ridge downstream.
	This asymmetry 
	of the air film layer
	along the direction of the wall motion is crucial for the generation of the lift. We will see in \S\ref{sec:model} that the essential features of the 3D profiles can be captured by a simple 2D model.
	
	We also 
	observed that the extension of the flat region of the air film increases with the drop size. For drops larger than typically 10\,mm, the drop shape is not stationary any more: it oscillates periodically until it eventually touches the surface. This suggests that an instability mechanism limits the size of the levitating drops to a value which is comparable with the capillary length.


\vspace{-7mm}
\section{Model}\label{sec:model}

	We now introduce a model for the levitating drop which both aims at clarifying the mechanism of the levitation and at explaining the typical thickness and shape of the air film separating the drop from the wall. We do not intend here to describe the 
	details of the 3D 
	shape of the air film. 
	Instead, we 
	rely on a simple 2D model that captures the essential features of the air film 
	and 
	allows
	for analytical predictions in the limit of both large and small drops.     
	
	We consider a stationary liquid drop levitating above a flat solid wall moving at constant velocity $\vect{V} = V\vect{e_x}$, parallel to its surface. The weight of the drop is sustained by an air film which is dragged under the drop. The shape of the drop results from a balance between surface tension, gravity and viscous stresses in the air, respectively involving the surface tension $\sigma$ of the liquid, the gravity $g$ and the density $\rho$ of the liquid, and the dynamic viscosity $\eta=1.81\,10^{-5}$\,Pa\,s of the air. As already mentioned, we consider a 2D drop, that is to say a liquid mass having a cylindrical shape which is invariant along the direction tangent to the solid surface and perpendicular to $\vect{V}$. We assume that the liquid in the drop is at rest (as observed in \S\ref{sec:expeobs}) and that inertial and compressibility effects in the air are negligible (the typical Reynolds number in the air film is $10^{-1}$). In that case, the drop shape only depends on
the capillary number $\ca =\eta V/\sigma$, 
the angle $\alpha$ between the direction normal to the wall and that of gravity (see figure \reff{schema}),
and the product $a\kappa_0$ of the capillary length 
$a=\sqrt{\sigma/\rho g}$ with the curvature $\kappa_0$ of the drop at its widest point.
	Considering the hydrostatic pressure field inside the drop, Laplace's equation relates the pressure $p$ in the air, at the interface, with the coordinates $\{x,z\}$ and the curvature $\kappa$ of the drop's interface according to
	\vspace{-1mm}
	\begin{equation}\label{eq:lap}
	\sigma(\kappa_0-\kappa) + \rho g\left[(z_0-z)\cos\alpha + (x_0-x)\sin\alpha\right] = p-p_0,
	\vspace{-1mm}
\end{equation}
	where $\{x_0,z_0\}$ and $p_0$ respectively stand for the coordinates and the reference pressure at the widest point of the drop (see figure \reff{schema}).
\begin{figure}
	\centerline{\includegraphics[width=.8\textwidth]{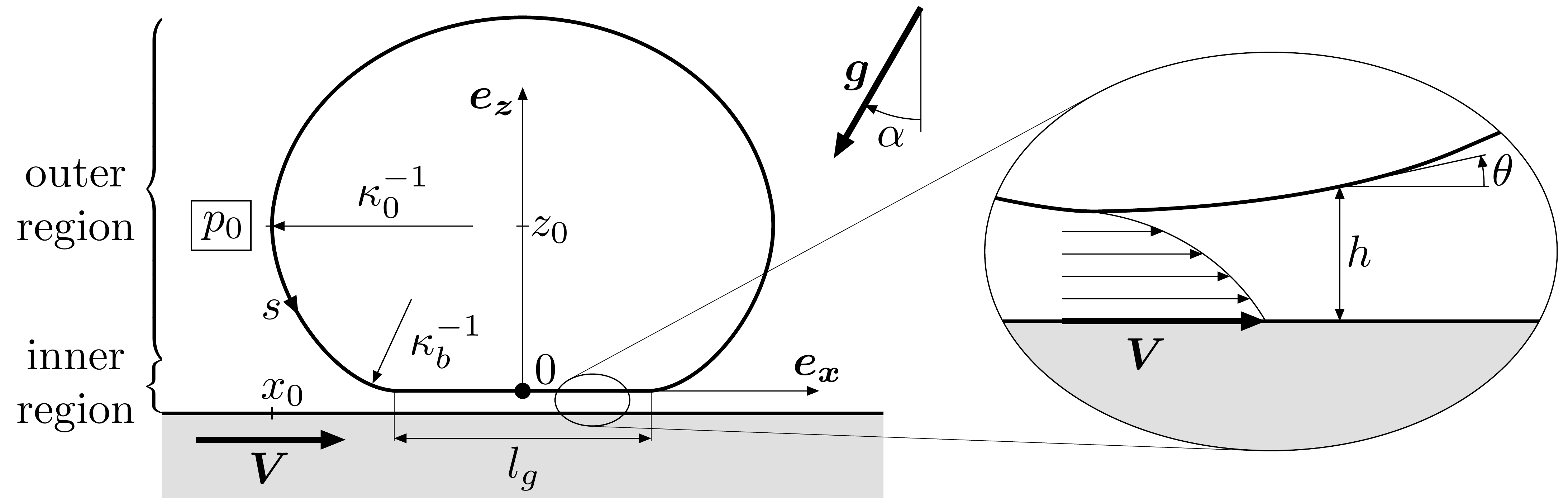}}
	\vspace{-3mm}
	\caption{Schematic of the levitating drop.}
	\label{fig:schema}
	\vspace{-5mm}
\end{figure}
 
 	We first consider 
	the limit of 
	 small capillary number,
	 which corresponds to our experimental conditions 
	 where $\ca \sim 10^{-3}$, and seek for asymptotic solutions. In that case, the influence of the viscous air flow on the drop shape is confined to the air film under the drop, as in the classical analysis by \cite{Landau42}. 
	 
	 On the one hand, in the outer region, i.e., far from the air film (see figure \reff{schema}), the profile of the drop is that of a sessile drop, that is to say prescribed by a balance between capillarity and gravity only. Letting $p=p_0$ in equation \refe{lap} yields, after integration, the coordinates $x_S$ and $z_S$ of the sessile drop's interface
\vspace{-1mm}
\begin{equation}\label{eq:xz}
	\frac{x_S}{a} =  \int_\pi^\theta \frac{\cos\theta}{\sqrt{a^2\kappa_0^2+2\cos(\theta+\alpha)}}\,\dd\theta, \qquad
	\frac{z_S}{a}  = \sqrt{a^2\kappa_0^2+2} - \sqrt{a^2\kappa_0^2+2\cos(\theta+\alpha)},
\vspace{-1mm}
\end{equation}
	where $\theta$ is the angle between the interface and the wall (see figure \reff{schema}). For the following, it is important to express the width $l_g$ of the sessile drop's base when $\alpha = 0$:
\vspace{-1mm}
\begin{equation}\label{eq:L}
	\frac{l_g}{a} = 2 \int_\pi^{-\pi} \frac{\cos{\theta}}{\sqrt{a^2\kappa_0^2+2\cos\theta}}\,\dd\theta = \frac{\pi}{a^3\kappa_0^3}+O\left(a^{-7}\kappa_0^{-7}\right).
\vspace{-1mm}
\end{equation}
	The approximation in the limit of small droplet sizes, i.e., when the drop shape is almost circular with radius $\kappa_0^{-1}$, is better understood by seeing that the drop weight $ \pi\rho g\kappa_0^{-2}$ has to balance the pressure force $l_g \sigma \kappa_0$ applying on the wall. It is accurate within 
	7\% for droplet radii smaller than $a$.
	
	On the other hand, in the inner region, i.e., that in contact with the air film, the profile of the drop results from a balance between the pressure due to the viscous flow in the air film and capillarity. Assuming a lubrication flow, the pressure in the film evolves according to
	\vspace{-4mm}
\begin{equation}\label{eq:pp}
	p' = -\frac{12\eta}{h^3}\left(q-\frac{Vh}{2}\right)
\end{equation}
	where $'$  denotes the derivative with respect to $x$, and $h$ and $q$ respectively stand for the thickness of the air film and the air flow rate per units of width in the film. Using the linearized form of the curvature $\kappa\simeq h''$, equations \refe{lap} and \refe{pp} combine into
	\vspace{-1mm}
\begin{equation}\label{eq:inner}
	a^2 h'''+\cos\alpha\,h'+\sin\alpha = 6 a^2\ca(2 q/V-h)/h^3.
\vspace{-1mm}
\end{equation}
	For the experimental conditions, $\ca\sim10^{-3}$ and $\sin\alpha\ll\ca(a/h)^2\sim10^2$ (which is consistent with the output of the model, see equation \refe{h0large}). Equation \refe{inner} can then be approximated by
\vspace{-2mm}
\begin{equation}\label{eq:hppp}
	h''' = 6 \ca(2q/V-h)/h^3.
\end{equation}
Introducing the dimensionless film thickness $H$ and coordinate $\xi$ defined as 
\vspace{-2mm}
\begin{equation}\label{eq:H}
	H(\xi) = h/h_0, \qquad\rm{with}\qquad \xi = x/l_\eta, \qquad l_\eta = h_0/(6\ca)^{1/3}, \qquad h_0 = 2q/V,
	\vspace{-1mm}
\end{equation}
	equation \refe{hppp}  adopts the well known dimensionless form
	\vspace{-2mm}
\begin{equation}\label{eq:Hppp}
	H''' = (1-H)/{H^3}.
\vspace{-2mm}
\end{equation}

\subsection{Large drops}\label{sec:largedrop}
\vspace{-1mm}

	As seen in \S \ref{sec:expeobs}, for ``large'' drops the film longitudinal profile exhibits a long and almost flat portion. We call this regime the large drop regime. It corresponds to the case when the length $l_g$ of the flat area at the bottom of the sessile drop is much larger than the length $l_\eta$ of the visco-capillary transition region introduced in equation \refe{H}.
	
	In the flat portion of the film $H'''\simeq 0$. One may thus write $H=1+\epsilon$ with $\epsilon\ll1$ and obtain
\vspace{-3mm}
\begin{equation}\label{eq:eps}
	\epsilon = A\ed^{-\xi}+B\cos\pen(\sqrt{3}\xi/2+\phi)\ed^{\xi/2},
\end{equation}
	after linearizing and solving equation \refe{Hppp}. As the length of the flat region separating the upstream from the downstream side of the drop can theoretically be made arbitrarily long, asymptotic expressions for each of them can be sought for separately, in the very same manner as in \cite{Bretherton61}. Note that this length can however be limited for stability reasons, as direct observations on large drop suggest. For the upstream portion, i.e., as $\xi \rightarrow -\infty$, equation \refe{Hppp} has to be solved with the asymptotic boundary conditions%
\vspace{-2mm}
\begin{equation}
	H(+\infty) = 1, \qquad H'(+\infty) = 0.
	\vspace{-1mm}
\end{equation}
	Using equation \refe{eps}, with the arbitrary constant $A=1$, and with $B=0$ corresponding to the previous boundary condition, numerical integration of equation \refe{Hppp} from $\xi = 20\gg1$ yields the asymptotic behavior $H''(\xi)\simeq0.643$ as $\xi\rightarrow-\infty$, as already found by \cite{Bretherton61}.
\begin{figure}
	\centerline{\includegraphics[width=.37\textwidth]{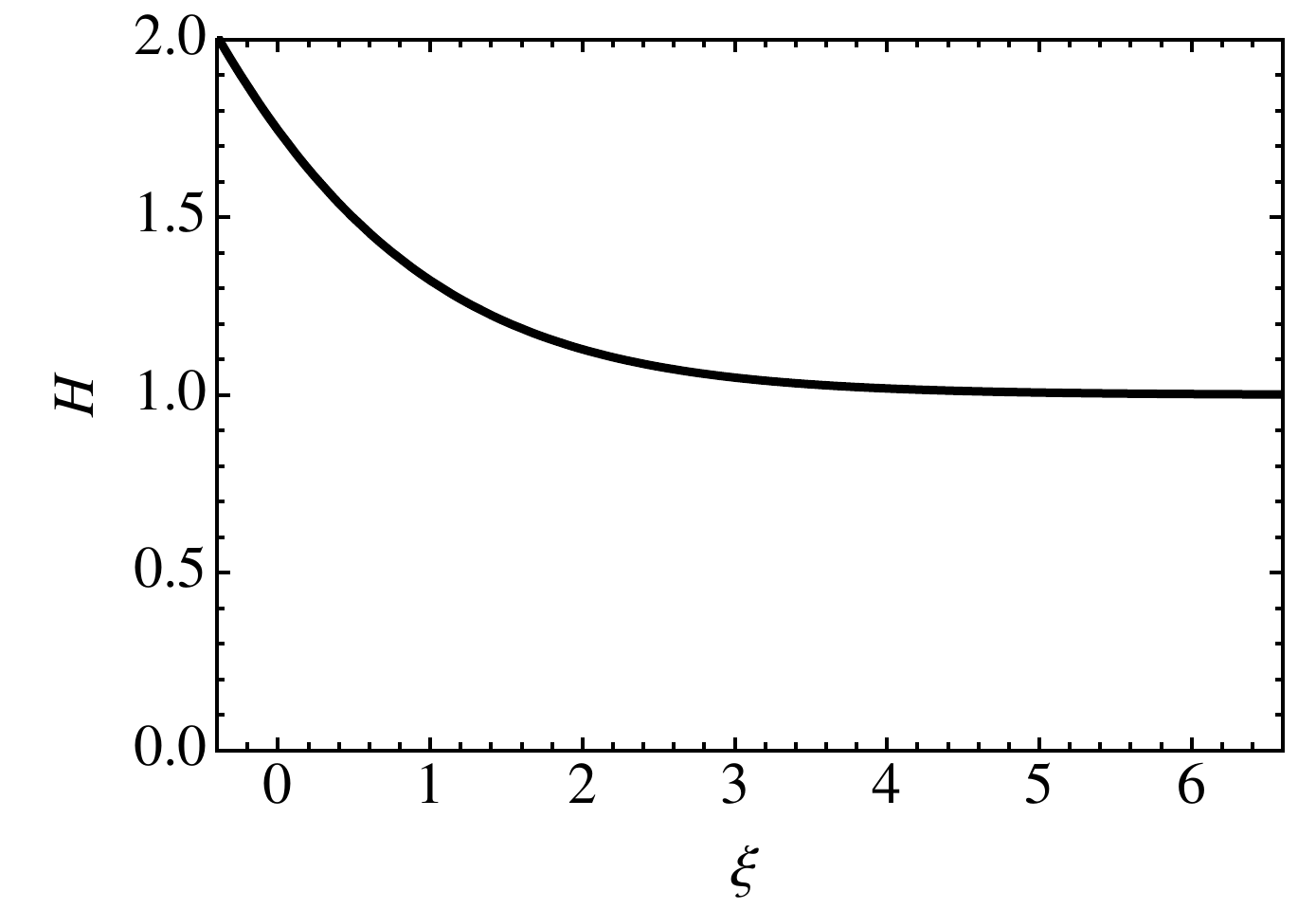}\includegraphics[width=.37\textwidth]{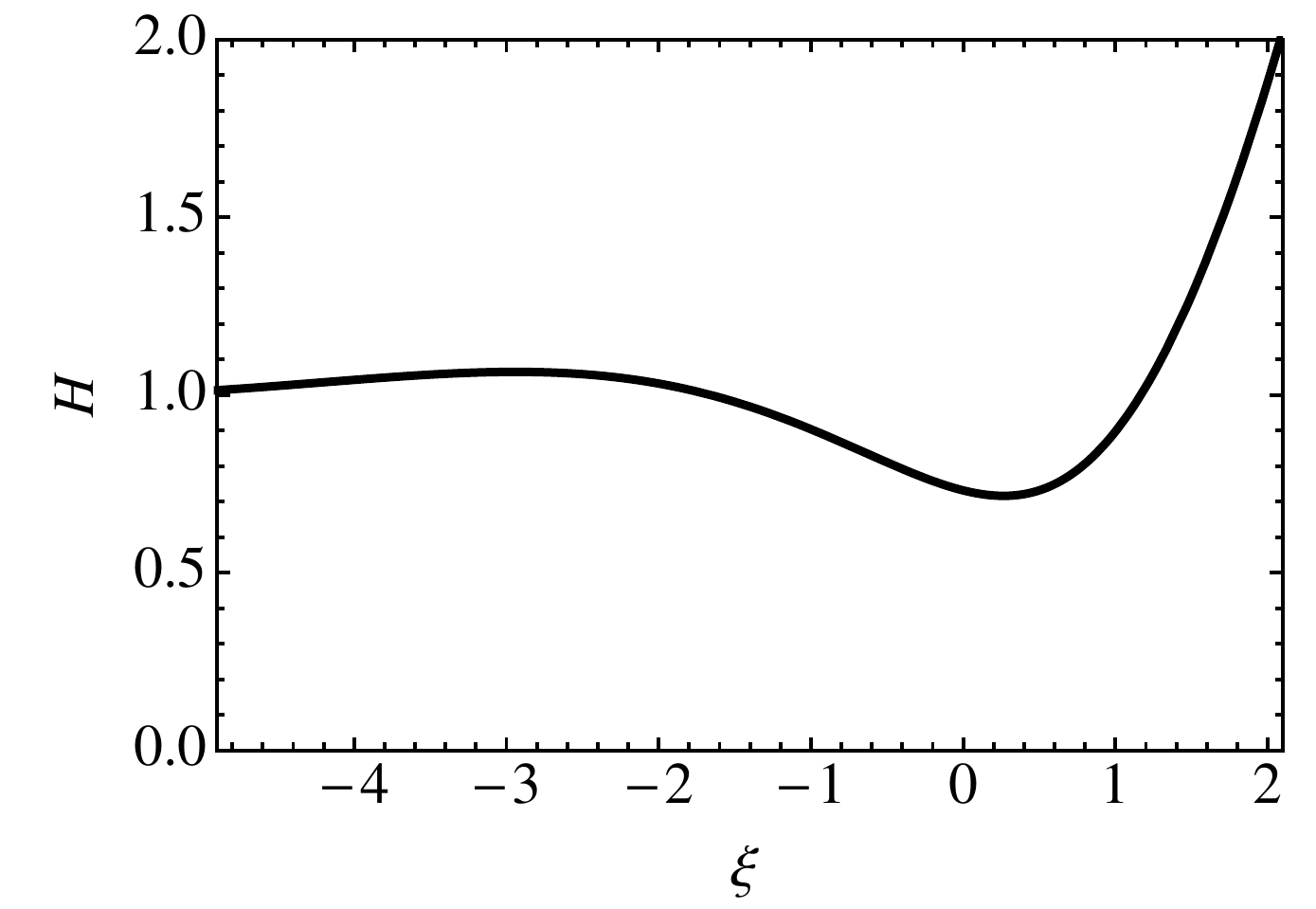}}
	\vspace{-4mm}
	\caption{Dimensionless upstream (a) and downstream (b) profiles of the air film thickness for a large drop.}
	\label{fig:proflarge}
	\vspace{-5mm}
\end{figure}
	
	For the downstream portion,  i.e., as $\xi \rightarrow +\infty$, equation (\ref{eq:Hppp}) has to be solved with the asymptotic boundary conditions
	\vspace{-2mm}
\begin{equation}\label{eq:Hinfty}
	H(-\infty) = 1, \qquad H'(-\infty) = 0, \qquad H''(+\infty)\simeq0.643,
	\vspace{-2mm}
	\end{equation}
	since, at the bottom of the drop, the same curvature has to be recovered as for the upstream side of the drop. 
	This is valid for the case $\alpha=0$.
	Equation \refe{eps} is now used with $A=0$ and the arbitrary constant $\phi=0$. $B$ is determined by a shooting method, so as to match the last boundary condition. This yields $B\simeq0.321$ and a master shape exhibiting a minimum $H_{min} \simeq 0.716$ and a first local maximum $H_{max} \simeq 1.065$ as shown on figure \reff{proflarge}. 
	
	By matching the asymptotic constant curvature $H''(+\infty)$ of boundary condition \refe{Hinfty} with the actual curvature $h'' = \kappa_b$ at the bottom of the sessile drop (see figure \reff{schema}), one obtains the air film thickness $h_0$ in the flat portion under the drop
	\vspace{-1mm}
\begin{equation}\label{eq:h0large}
	h_0 \simeq 2.123\,\ca^{2/3}\kappa_b^{-1}.
\vspace{-2mm}
\end{equation}
	Note that $\kappa_b$ is related to $\kappa_0$ through the shape of the sessile drop: for a 2D drop, equation \refe{xz} yields $\kappa_b = \sqrt{\kappa_0^2+2/a^2}$. In the asymptotic regime, the dimple in the air film, on the downstream profile, can be characterized by the minimal thickness $h_{min}\simeq 0.716\,h_0$, the first local maximal thickness $h_{max}\simeq1.065\,h_0$ and the horizontal spacing $\lambda\simeq4.238\,\ca^{1/3}\kappa_b^{-1}$ between them.
	Note also that, in the case $\alpha>0$, the downstream profile would only be slightly modified, because the drop curvature $\kappa_b$ would be smaller than that of the upstream side of the drop, but this would hardly modify $h_0$, as, for small \ca, $h'' = 0$ in equation \refe{inner} yields
	\vspace{-1mm}
\begin{equation}
	\frac{h}{h_0} \simeq 1-\frac{\sin\alpha}{0.643^2 (a\kappa_b)^{2}}(6\ca)^{1/3} \simeq 1-0.751\frac{\sin\alpha}{(a\kappa_b)^{2}}\ca^{1/3},
	\vspace{-1mm}
\end{equation}
	which is very close to 1 for the present experimental conditions.
	\vspace{-2mm}
	
\subsection{Small drops}\label{sec:smalldrop}
\vspace{-1mm}
	For ``small'' drops, the shape of the air film is radically different. The upstream and downstream visco-capillary regions merge, and the drop adopts an almost circular shape while the air film profile become close to parabolic. This corresponds to the case where $l_g<l_\eta$. One therefore naturally seeks for a film thickness that is a perturbation of a parabolic profile $H_0(\xi)$, with an {\it a priori} unknown dimensionless curvature $K$, corresponding to an undeformed drop
\vspace{-2mm}
\begin{equation}
	H = H_0 + \epsilon = C + K\xi^2/2 + \epsilon,
\end{equation}
	where $C$ is an unknown constant and 
	$\epsilon \ll H_0$
	is the condition that will be checked {\it a posteriori}. Equation \refe{Hppp} then becomes
\vspace{-1mm}
\begin{equation}
	\epsilon''' = (1-H_0)/H_0^3 + O(\epsilon),
	\vspace{-2mm}
\end{equation}
	which has a single real solution
	\vspace{-1mm}
\begin{equation}
	\epsilon = - \left(2/(3K)\right)^{3/2}\arctan\left(\sqrt{2 K/3}\xi\right),\qquad\qquad\rm{for}\qquad\qquad C=3/4.
\vspace{-1mm}
\end{equation}
	This solution is
	plotted in figure \reff{profsmall} for $K=1$. 
\begin{figure}
	\centerline{\includegraphics[width=.37\textwidth]{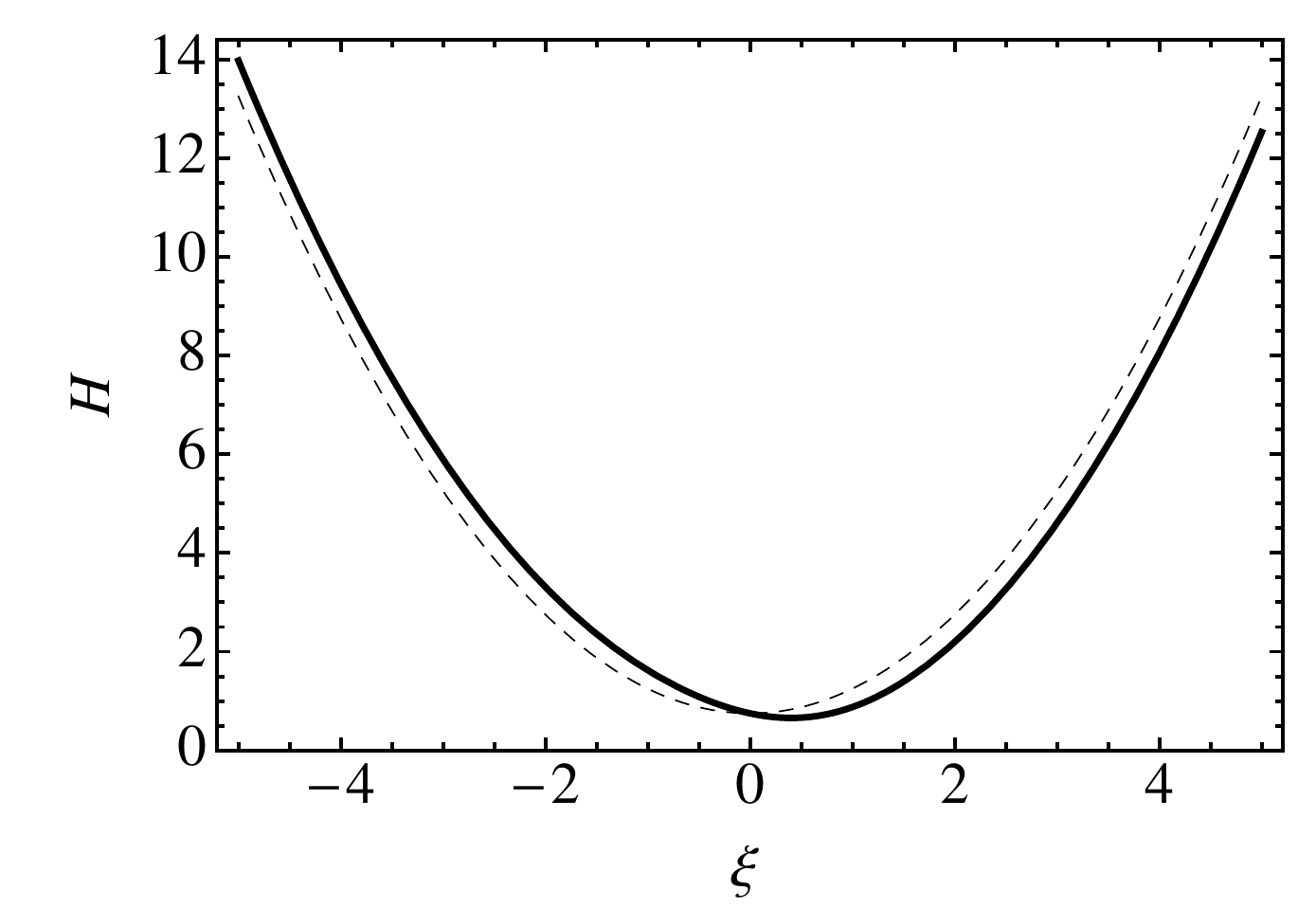}\includegraphics[width=.37\textwidth]{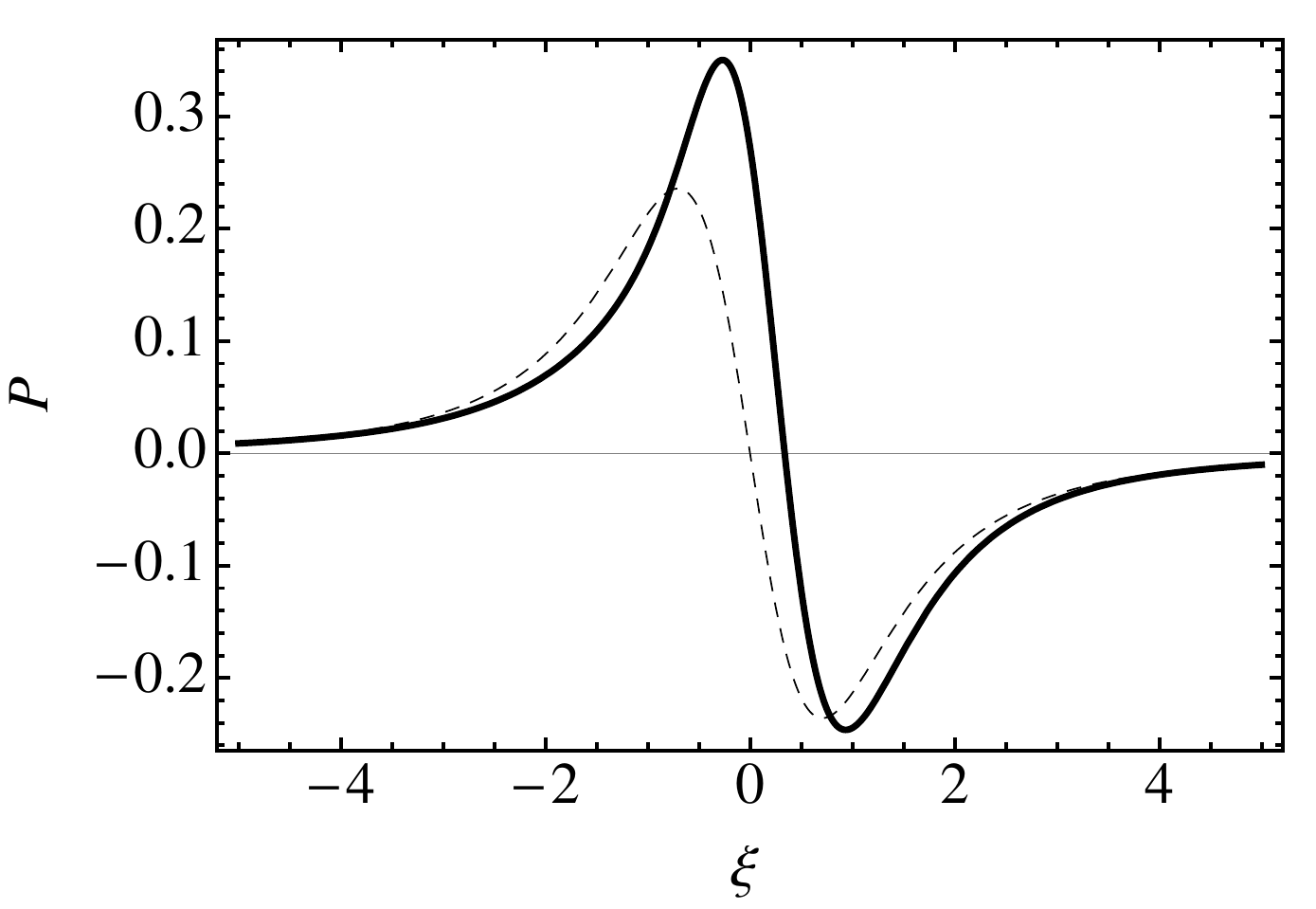}}
	\vspace{-4mm}
	\caption{Dimensionless air film thickness (a) and pressure in the film (b) for a ``small'' drop (plotted for $K=1$). The dashed and plain lines respectively correspond to the undeformed parabolic profile $H_0$ and the deformed profile $H_0+\epsilon$.}
	\label{fig:profsmall}
		\vspace{-5mm}
\end{figure}

	In order to determine both $K$ and $h_0$, two matching conditions have now to be found. The first one is the matching of the curvatures $H''(\pm\infty)$ with $\kappa_b$, as in \S \ref{sec:largedrop}. The only other relevant matching possibility is global: the 
	resultant of the 
	pressure has to balance the drop's weight (note that this was automatically verified in \S\ref{sec:largedrop} by connecting a flat film portion with one having a uniform curvature $\kappa_b$). One therefore needs to determine the pressure field in the air film. This field can be thought of as the sum of an odd contribution 
$P_0$ that results from the parabolic shape $H_0$
and a small symmetric contribution $P_1$ that results from the small odd deformation $\epsilon$, 
\vspace{-1mm}
\begin{equation}\label{eq:P0}
	P \simeq P_0 + P_1 = - \int_{-\infty}^x\frac{H_0-1}{H_0^3}\,\dd\xi + P_1 = -\frac{16\xi}{3(3 + 2 K \xi^2)^2}+ P_1,
\vspace{-2mm}
\end{equation}
	where the dimensionless Laplace's law $P'=-H'''$ has been used. $P_1$ is then determined from the first order term in $\epsilon$ in equation \refe{Hppp} with the boundary condition $P_1(\pm\infty)=0$ yielding
\vspace{-3mm}
\begin{equation}\begin{aligned}
	P_1 &= -\int\frac{2 H_0-3}{H_0^4}\epsilon\,\dd\xi = \frac{2^3}{3^4K^2}\bigg[\pi^2-4\arctan^2\left(\sqrt{\frac{2 K}{3}}\xi\right)\\
	&-24\frac{(8+7K\xi^2+2K^2\xi^4)+\sqrt{\frac{2K}{3}}\xi(27+16K\xi^2+4K^2\xi^4)\arctan\left(\sqrt{\frac{2 K}{3}}\xi\right)}{(3+2K\xi^2)^3}\bigg].
	\vspace{-2mm}
\end{aligned}\end{equation}
	Since $P_0$ is odd, it does not contribute to the total pressure, and $P_1$, in its dimensionless form, is then
\vspace{-3mm}
\begin{equation}\label{eq:F_Z}
	\frac{f_z}{\sigma\ca^{1/3}} = F_Z = \int_{-\infty}^{\infty} P_1\cos\theta\,\dd\xi \simeq \int_{-\infty}^{\infty} P_1\,\dd\xi = \frac{\pi}{9}\left(\frac{2}{3K}\right)^{5/2}.
\vspace{-2mm}
\end{equation}
	Note that equation \refe{F_Z} indeed represents the dominant contribution to the vertical force, since the zero order term in $\int\tau\sin\theta\,\dd x$ is exactly {\it nil}, while the first order term is of order $\sigma\ca^{2/3}K^{-5/2}$, which will be shown to be much smaller than $\sigma\ca^{1/3}F_Z$ for small capillary numbers in equation \refe{hmin}.

	Finally, the two matching conditions $\kappa_b = h_0  K/l_\eta^2=(6\ca)^{2/3}K/h_0$, for the interface curvature, and $\pi (a\kappa_0)^{-2}\cos\alpha =  6\ca(l_\eta/h_0)^2 F_Z = (6\ca)^{1/3}F_Z$, for the weight of the drop, yield the actual minimal air film thickness
	\vspace{-1mm}
\begin{equation}\label{eq:hmin}
	h_{min} = \frac{3}{4} h_0 = 2^{-1/5}\ca^{4/5}\left(\frac{a^2\kappa_0^2}{\cos\alpha}\right)^{2/5}\kappa_b^{-1} \simeq 0.871\,\ca^{4/5}\left(\frac{a^2\kappa_0^2}{\cos\alpha}\right)^{2/5}\kappa_b^{-1}.
\vspace{-2mm}
\end{equation}

\subsection{Air drag}\label{sec:drag}

	In the experiment, the position of the drop in the rotating cylinder is determined by the ratio of its weight to the air drag. Beside the ``body'' drag due to the main air flow around the drop, the air in the lubrication film flow under the drop generates an additional drag that can be dominant
	This latter drag {\it a priori} has two contributions: One due to the shear stress at the drop interface
\vspace{-1mm}
\begin{equation}\label{eq:tau}
	\int\tau\cos\theta\dd x = -\int \eta u_z|_h \cos\theta\,\dd x = \sigma (6\ca)^{2/3}\int\frac{3-2H}{6H^2}\cos\theta\,\dd\xi,
\vspace{-1mm}
\end{equation}	
	and the other due to the pressure 
\vspace{-1mm}
\begin{equation}\label{eq:p}
	-\int p\sin\theta\dd x = -\sigma \int (\kappa_b-\kappa) \sin\theta\,\dd x = \sigma (6\ca)^{1/3}\int\int\frac{1-H}{H^3}\sin\theta\,\dd\xi\dd\xi.
\vspace{-1mm}
\end{equation}

	For large drops, the shear stress by definition applies over a length $l_g$ that is much larger than the typical length $l_\eta$ and therefore, even for small capillary numbers, both drag terms are important. The pressure term \refe{p} in the flat portion of the film vanishes, as $\sin\theta=0$, the contribution of the upstream and downstream transition regions is numerically estimated from the profiles in figure \reff{proflarge}, while the dimensionless drag term is simply $(1/6)l_g/l_\eta$, since $H=1$ in the flat region. This yields
\vspace{-1mm}
\begin{equation}
	f_x = \sigma[1.224\,(6\ca)^{1/3}+l_g(6\ca)^{2/3}/6{l_\eta}] \simeq \sigma\left(2.22+0.471\,l_g\kappa_b\right)\ca^{1/3}
\vspace{-2mm}
\end{equation}	
	
	For small drops, on the other hand, the shear stress applies over a typical length $l_\eta$ and therefore, for small capillary numbers, the pressure term of the air drag is dominant. Using equation \refe{p}, with the zero order pressure field $P_0$ from \refe{P0}, yields the resultant force tangential to the wall%
\vspace{-3mm}
\begin{equation}\begin{aligned}
	\frac{f_x}{\sigma(6\ca)^{1/3}} = - \int_{-\infty}^{\infty} P_0\sin\theta\,\dd\xi = \frac{8K\left[2\sqrt{3} \arccos\sqrt{\frac{3K}{2}}-3\sqrt{(2-3K)K}\right]}{9\left[(2-3K)K\right]^{3/2}} = \frac{8}{9K} + O(K^{-2}),
\end{aligned}
	\vspace{-1mm}
	\end{equation}

		which also expresses as
	\vspace{-3mm}
\begin{equation}
	f_x = 2^{11/5}\sigma\ca^{1/5}\left(\frac{a^2\kappa_0^2}{\cos\alpha}\right)^{-2/5}\simeq 4.594\,\sigma\ca^{1/5}\left(\frac{a^2\kappa_0^2}{\cos\alpha}\right)^{-2/5}.
	\vspace{-2mm}
\end{equation}

\vspace{-2mm}
\subsection{Numerical integration}	
\vspace{-1mm}	
	Parallel to the asymptotic approach used in \S\ref{sec:largedrop} to \S\ref{sec:drag}, we also numerically integrated the nonlinear form of equation \refe{inner}, i.e.%
	\vspace{-1mm}
\begin{equation}\label{eq:num}
	\partial_{ss} \theta + \frac{\sin(\theta+\alpha)}{a^2} + 6\ca\cos\theta\frac{h-2q}{h^3} = 0,\qquad
	\partial_{s} h = \sin\theta,\qquad
	\partial_{s} x = \cos\theta,
	\vspace{-1mm}
\end{equation}
	where $s$ stands for the curvilinear coordinates along the drop's interface profile, with the following initial conditions%
	\vspace{-1mm}
\begin{equation}
	\theta(0) = -\pi/2,\hspace{3mm} \partial_{s} \theta(0) = \kappa_0,\hspace{3mm} h(0) =  z_S(-\pi/2)+\Delta h,\hspace{3mm} x(0) = x_S(-\pi/2).
	\vspace{-1mm}
\end{equation}	
	A shooting method was then used, with the {\it a priori} unknown flow rate $q$ and vertical shift of the drop $\Delta h$ (with respect to the sessile configuration) as adjustable parameters, so as to match the initial slope, curvature and sessile drop width in $\theta = \pi/2$.
	
	This approach permits to study configurations where the capillary number is not small. We 
	use it here to illustrate the cross-over between the large and the small drop regimes 
	that we discussed previously,
	and also for the case when $\alpha \neq 0$. Figure \reff{profnum} shows three numerical profiles 
	for three different drop sizes.
	These profiles illustrate the formation and the evolution of a plateau in the air film thickness
	as the drop size increases.	
\begin{figure}
	\centerline{\includegraphics[width=.4\textwidth]{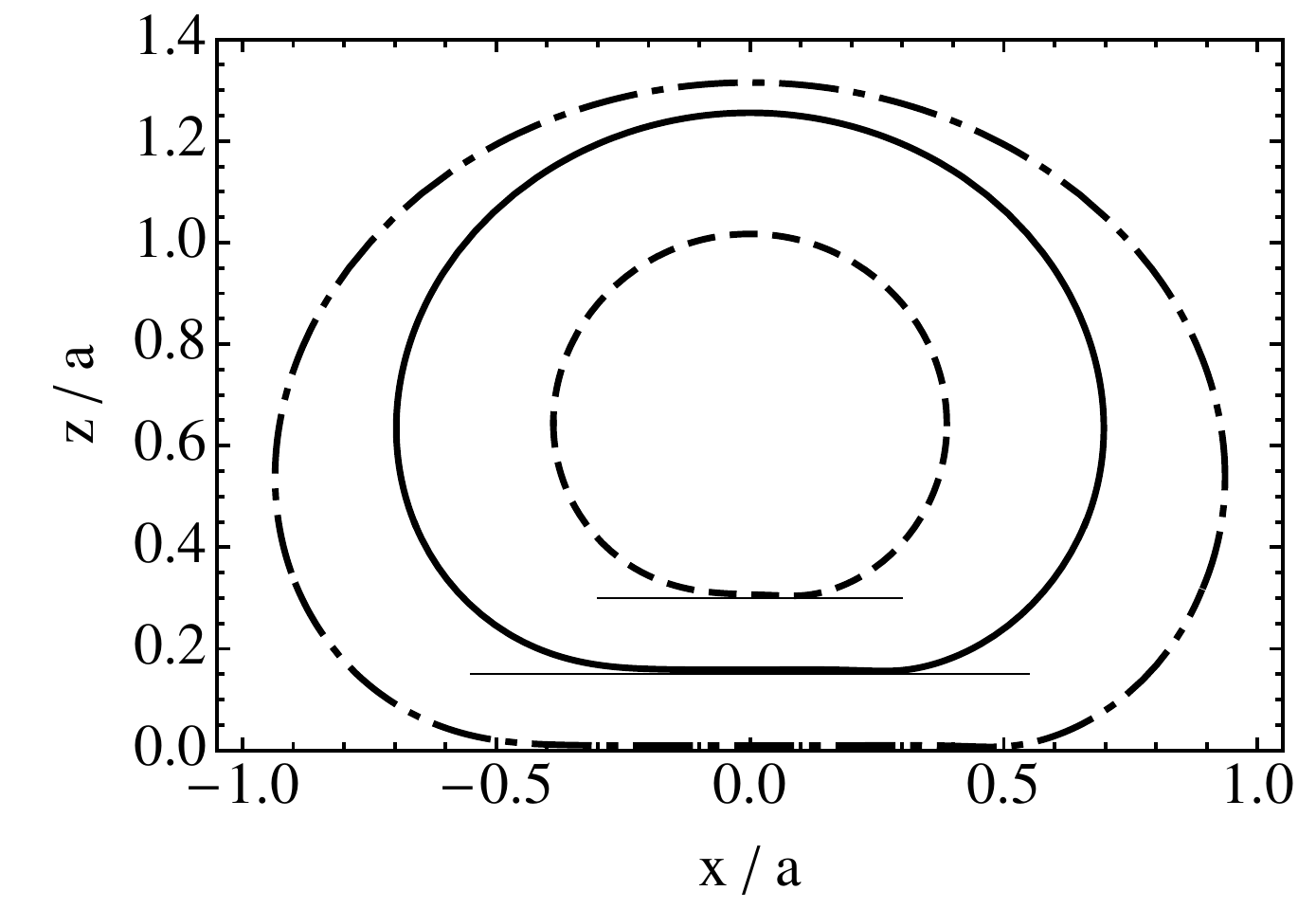}\includegraphics[width=.4\textwidth]{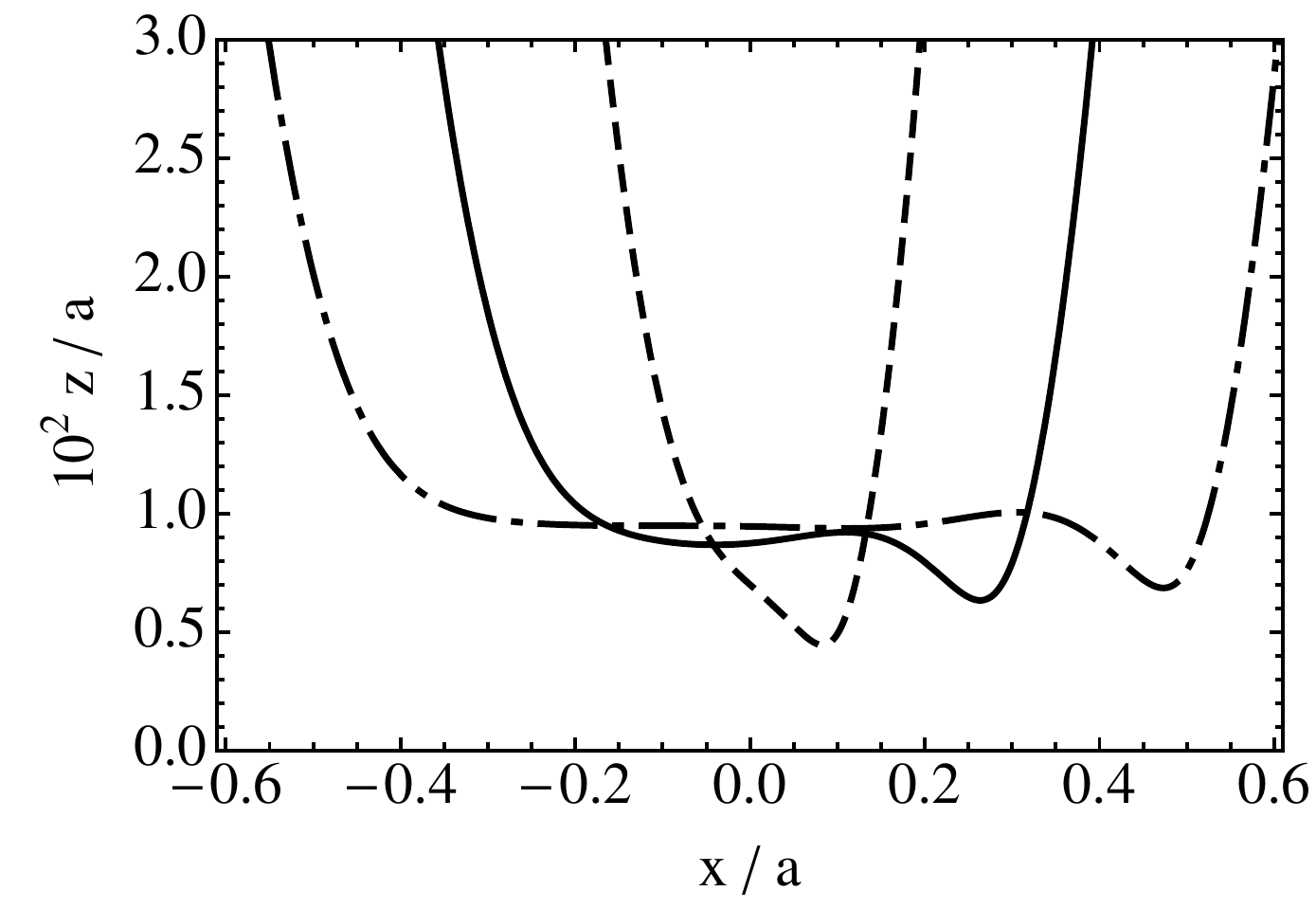}}
	\vspace{-4mm}
	\caption{Drop shapes (a) and air film thicknesses (b) obtained by numerical integration of \refe{num}. $\ca = 10^{-3}$, the wall is horizontal ($\alpha =0$), and the curvatures at the equator are $a\kappa_0 = 2.5$ (dashed line), 1.4 (plain line) and 0.9 (dashed-dotted line). The first and second profiles in (a) are vertically shifted by 0.3 and 0.15\,$a$ respectively.}
	\label{fig:profnum}
	\vspace{-3mm}
\end{figure}

\vspace{-6mm}
\section{Comparison}\label{sec:comp}

	We now turn to the comparison of the predictions of \S\ref{sec:model} with the air film thickness 
	we measured.
	In order to compare the 3D shape of the air film with the predictions of the 2D model, we focus on the profile in the plane of symmetry of the drop, i.e. parallel to the direction $\vect{e_x}$ of the wall motion. As we will see, it concentrates all the important characteristics of the air film necessary to explain the levitation of a drop.   

	Figure \reff{comp} shows three air film thickness profiles, for different drop sizes with similar wall velocities. The parameters of the experiments are listed in table \ref{tab:comp}. As already mentioned in \S\ref{sec:expeobs}, the air films are typically a few micrometer thick. Their shapes however strongly depend on the drop size, as predicted in \S\ref{sec:model}. For the smallest drop, with apparent diameter $D$ = 1.10\,mm = 0.74\,$a$ (figure \reff{comp}(a)), the thickness decreases almost continuously with $x$, until a minimal value is reached at the upstream side of the air film. The plateau predicted for large drops is hardly observed around $x\simeq0.4\,$mm. This is fully consistent with our model: here $l_g/l_\eta = 1.85$ is comparable to one, and therefore the capillary-viscous transition regions cover most of the drop's base. As the drop diameter is increased (figure \reff{comp}(b-c)), a plateau appears in the film thickness profile.
	It separates the two transition regions which exhibits similar shapes to those predicted in \S\ref{sec:largedrop}: the upstream profile is monotonic, while the downstream profile oscillates and shows a marked minimum just at the end of the drop's base.
\begin{figure}
	\centerline{\includegraphics[width=.3\textwidth]{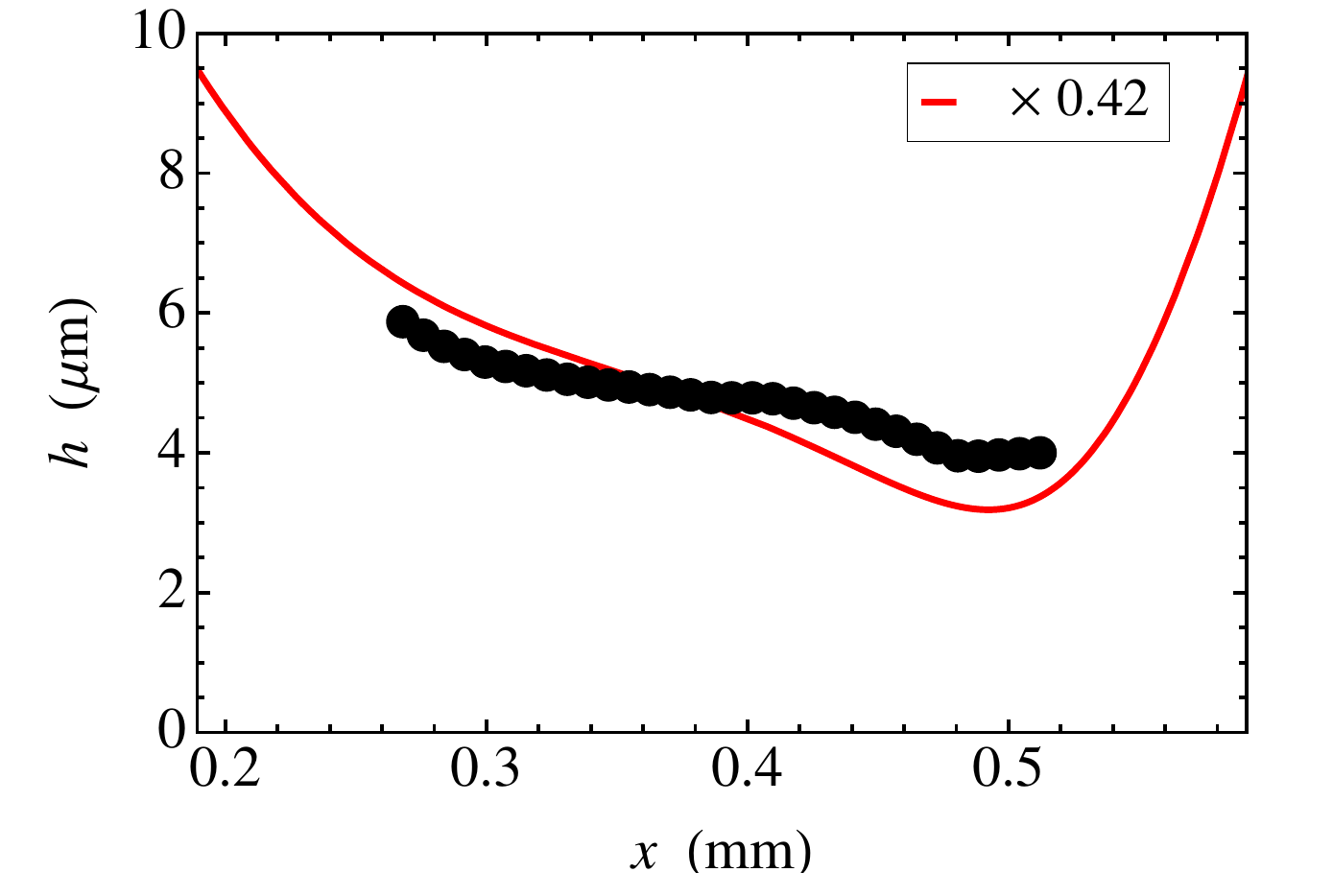}\includegraphics[width=.3\textwidth]{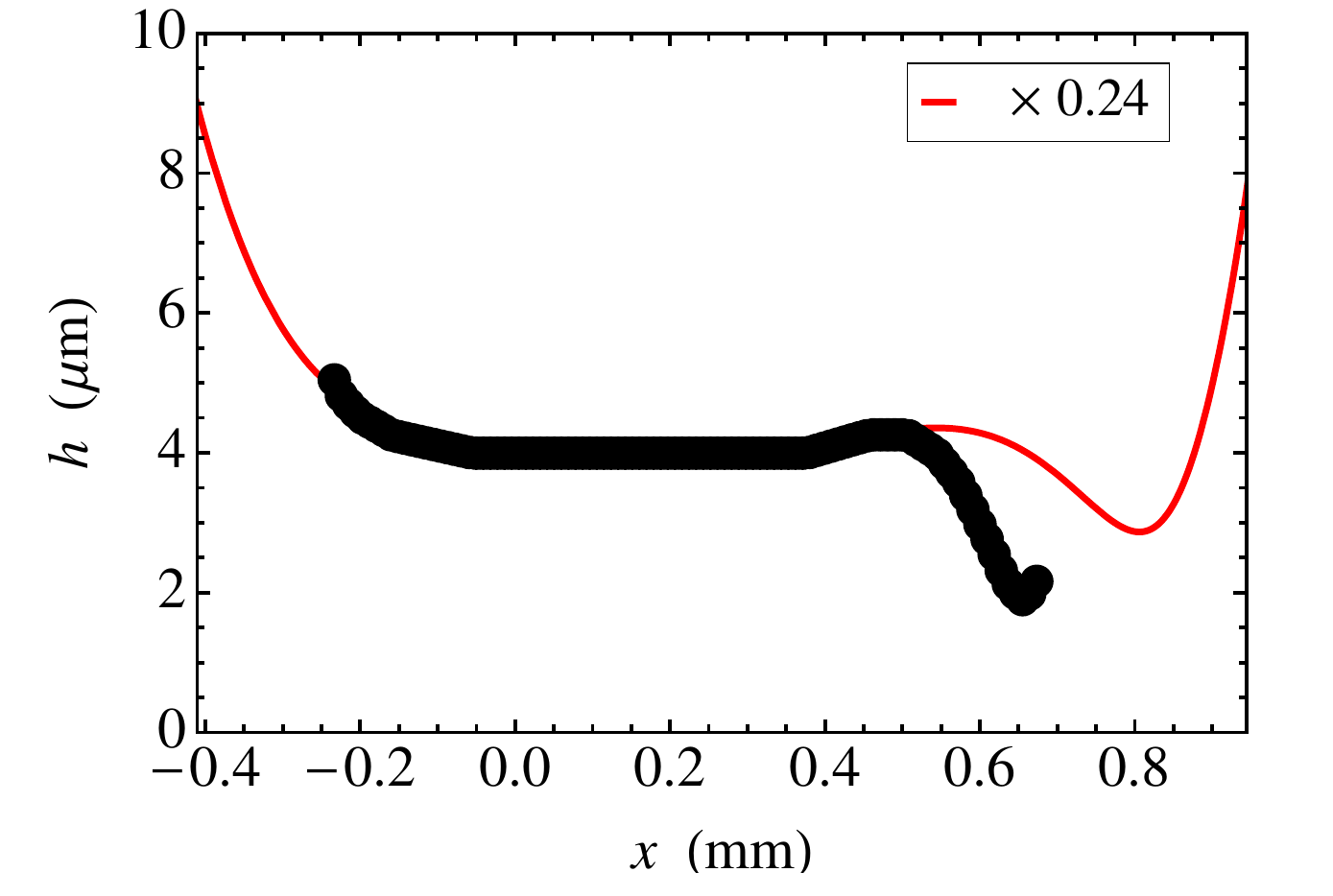}\includegraphics[width=.3\textwidth]{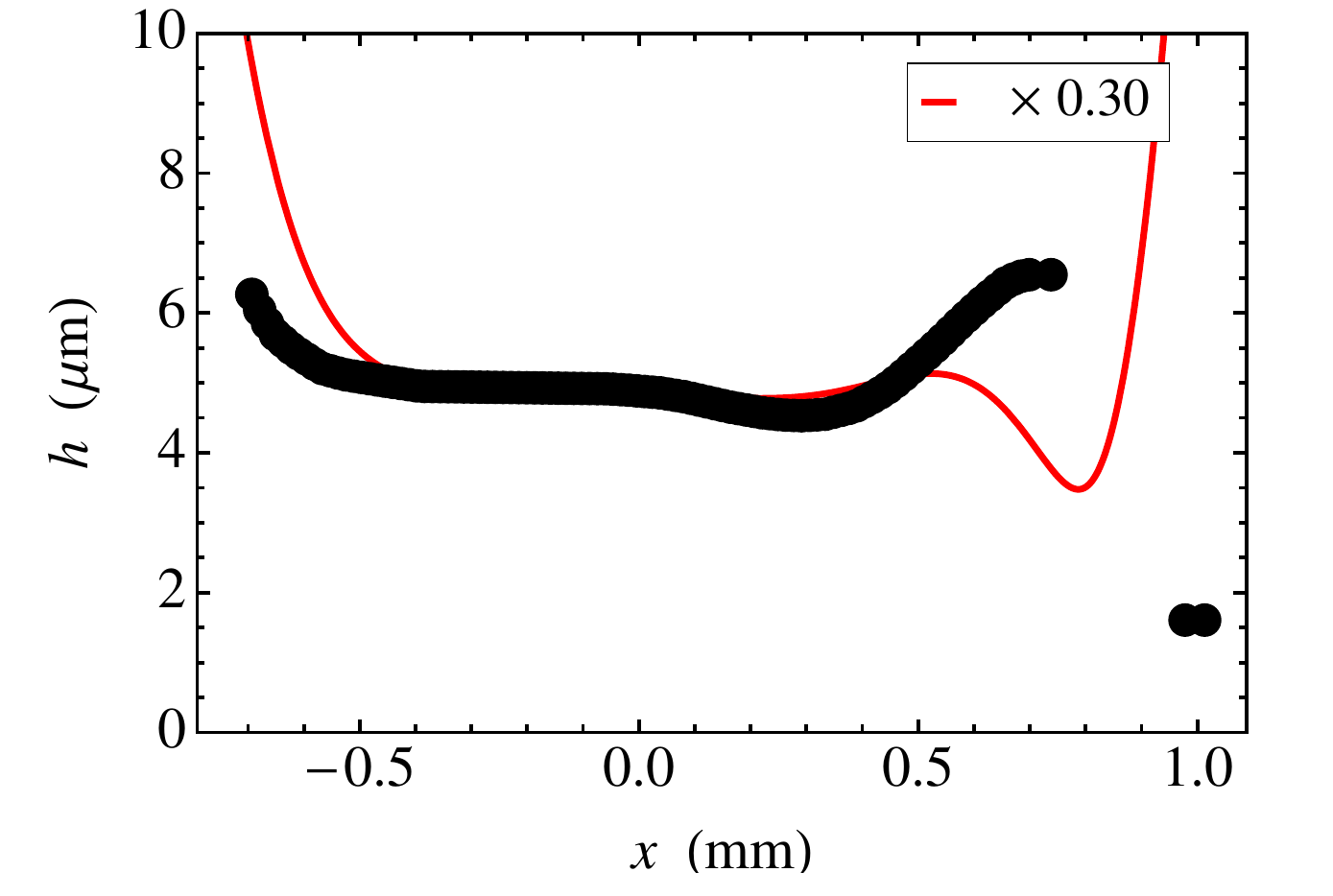}}
	\vspace{-4mm}
	\caption{Comparison of the experimental profiles of the air film, measured in the plane of symmetry (circles), with the numerical two dimensional profiles (plain lines) computed with gravity tilted by $\alpha$. (a), (b) and (c) respectively corresponds to the drops \#1, \#2 and \#3 whose parameters are listed in table \ref{tab:comp}. The thickness from the numerical profiles are respectively multiplied by 0.42, 0.24 and 0.30 to show the similarities in the shapes.}
	\label{fig:comp}
		\vspace{-5mm}
\end{figure}
\begin{table}\begin{center}
	\vspace{-3mm}
	\center\begin{tabular}{ccccccccccc}
		Drop \# &$V$	&$D$	&$\alpha$		&$h$	&$10^{3}\ca$  &$a$	&$l_g/l_\eta$  &$\kappa_b^{-1}$ &$h_0$&$\alpha_{eq}$\\[0pt]
		&(m$\cdotp{}$s$^{-1})$  &(mm) &($^{\circ}$)  &($\mu$m)  &(—)  &(mm)  &(—)  &(mm)  &($\mu$m)  &($^{\circ}$)\\
		1	&1.43	&1.10	&40.5	&4.8		&1.24	&1.49	&1.85	&0.250	&~6.3	&53.9\\
		2	&1.65	&2.32	&13.5	&4.0		&1.44	&1.49	&2.90	&0.425	&11.5	&~8.6\\	
		3	&1.43	&2.77	&~3.3	&5.0		&1.24	&1.49	&3.22	&0.467	&11.5	&11.6
		\vspace{-4mm} 
   	\end{tabular}
 	\caption{Experimental parameters and predictions corresponding to the three profiles shown in figure \reff{comp}. $a=1.49\,$mm. $\kappa_b$ is computed from $D$ and $a$, assuming a 3D axi-symmetric drop. $h_0$ is computed from \refe{h0large} corresponding to the large drop limit. The equivalent inclination angle is $\alpha_{eq} = \rm{asin}(f_x/f_z)$.
	}
	\label{tab:comp}
\end{center}
\vspace{-2mm}
\end{table}
	
	The fact that both the general shape and the main characteristics of the air film are correctly captured by the 2D model, should not hide that, although the absolute thicknesses are of the same order as in the experiments, they still differ by a substantial factor: the experimental thickness is systematically smaller than that from the model, whether one considers the analytical results, in table \ref{tab:comp} (factors of 1.3 to 2.9 when considering the curvature $\kappa_b$ of a 3D drop with apparent diameter $D$), or the numerical integration, in figure \reff{comp} (factors 2.4 to 4.2, as $\kappa_b$ corresponds to a 2D drop with the same $D$). 
	The assumptions about the  air flow in the film could, {\it a priori}, explain some of this difference. 
	Indeed, due to the motion in the liquids, the velocity in the air film is not exactly $V$ at the liquid layer's surface, neither exactly $0$ at the drop's interface. 
	This results in a slightly smaller lift force than predicted for the same film thickness, i.e., a smaller thickness to levitate the drop. This could explain why a low viscosity drop is more difficult to levitate: it has a comparatively thinner air film underneath. 
	However, as already stated in \S\ref{sec:expeobs}, for the fairly high viscosity we used, we could experimentally check that the no-slip condition is a very good approximation. We therefore think the discrepancy is mainly due to the difference between the two and 3D configurations. Indeed, in the former configuration the air flow rate is conserved along $x$, while in the latter some of the air, which is dragged under the drop, escapes on the sides resulting in a smaller value of $h_0$ necessary to built up the same lubrication pressure.
	
	We moreover think that, although the pre-factors are different for the real 3D drops, the scaling laws in \refe{h0large} and \refe{hmin} 
	are still valid and provide important predictions for the film thickness and air drag due to the lubrication flow.

	It is also important to realize that any model assuming lubrication theory in the air film, including ours, 
should not be
able, alone, 
to account for the existence of a velocity threshold for the levitation of a drop. As equations \refe{h0large} and \refe{hmin} show, for smaller and smaller capillary numbers, the film thickness is simply expected to decrease (while the accuracy of the model keeps increasing). We therefore think that the velocity threshold observed in the experiment is rather due to the existence of a minimum in the air film thickness that one can practically sustain in the presence of noise and wall roughness, than a strict minimum dictated by considerations on a smooth geometry. This is 
	consistent
	with the facilitated levitation observed over a liquid surface.%
%
%
\vspace{-6mm}
\section{Conclusion and extension}\label{sec:concl}
\vspace{-1mm}
	The levitation of a drop over a moving surface has been studied 
	using a
	simple and original setup, a rotating cylinder.
	We observed that the levitation was possible for different drop sizes above a critical surface velocity that increased with drop size. The levitation was experimentally facilitated by covering the cylinder's surface with a thin liquid layer and by using a liquid with intermediate viscosity $\nu\sim50\,$cSt. Both smaller and larger viscosities resulted in lower stability and levitation time of the drop. Interferometric measurements yielded the shape and the absolute thickness of the air film 
	underneath the drop. 
	It consists of a flat region, extending with increasing drop size, and surrounded by a ridge of minimal thickness of the air film, on the downstream and lateral sides. As we showed with a simplified 2D model, this shape is a consequence of the deformation of the drop's interface by the pressure due to lubrication flow in the air film. The pressure built-up is responsible for the lift force and explains the levitation of the drop. The 2D model also provides predictions for the air film shape, its absolute thickness and the air drag due to the lubrication flow under the drop, for two asymptotic regimes whose scalings are expected to be valid in the real 3D configuration: large drops, when the bottom of the drop is essentially flat, and small drops, when it is almost spherical.
	Let us conclude by mentioning some further investigations about levitating drops that the present study naturally asks for.  Measurements of the film thickness and air drag need to be extended to check the validity of the asymptotic predictions we proposed. The precise influence of the liquid viscosity is still not known, so is the criteria for the practical velocity threshold for levitation. Finally, the steady wake of the drop and the interaction between levitating drops would also worthwhile being studied. The experimental system we proposed provides an ideal configurations for those investigations.
\tb{We greatly thank Deflef Lohse for the wonderful environment and the setup to conduct this research.}

\vspace{-6mm}		
\bibliographystyle{jfm}

\end{document}